# Two-Fluid Model for Granular Superconductors

C. A. M. dos Santos*, C. J. V. Oliveira, M. S. da Luz, A. D. Bortolozo,
M. J. R. Sandim, and A. J. S. Machado
*Grupo de Supercondutividade, Departamento de Engenharia de Materiais, FAENQUIL, Caixa Postal 116,
12600-970, Lorena-SP, Brazil.*

A two-fluid model is proposed to describe the transport properties of granular superconductors. Using the resistively shunted junction model and some aspects of the two-level system theory, a statistical model is developed which takes into account the ratio between the number of normal and superconducting electrons carrying the applied current. The theoretical model reveals excellent agreement when compared to transport properties of four high-$T_C$ superconductors. The results suggest that the two-fluid model is independent of the sample composition, critical temperature and whether the superconducting compound is electron or hole-doped.

## I. INTRODUCTION

During the last 40 years, much attention has been devoted to describing transport properties of superconductors. The first description was based on the motion of Abrikosov vortices.[1-6] Nowadays flux-creep and flux-flow mechanisms are well accepted to describe, respectively, non-linear and linear parts of current-voltage ($I-V$) characteristic curves of homogeneous superconductors.[5-7] In the flux-flow regime, the Lorentz force on an Abrikosov vortex is much higher than the pinning force resulting in vortex motion on a viscous medium under the influence of Magnus force due to supercurrents circulating around the vortex core.[5,6] In such a model, flux-flow resistance is given by $R_f = R_N H / H_{C2}$ at low temperatures, where $H$ is the applied magnetic field, $H_{C2}$ is the upper critical field and $R_N$ is the normal state resistance. Dissipation in this regime is attributed to the relative motion of normal electrons inside the vortex core.[1-6] Furthermore, $I-V$ characteristic curves reveal linear dependencies with the applied current which are strongly magnetic field dependent (see, for example, reference [5]).

After the prediction of the Josephson effect,[8] several authors have taken into account aspects of thermal fluctuations or phase slippage,[9-15] two-fluid theory,[16] and the resistively shunted junction (RSJ) model[17-20] in order to describe transport properties of superconductors. In those models, shunt resistances are always present and normal current can flow parallel to the supercurrents in the dissipative regime ($V \neq 0$).[9-20] With the discovery of high-$T_C$ superconductors, several researchers have integrated aspects of weak coupling or granularity in order to understand electrical properties of this new class of superconductors.[21] Granularity in those superconductors is generally manifested through a double superconducting transition,[13-15,22-25] characterized by two superconducting critical temperatures labeled $T_{Ci}$ and $T_{Cj}$.[22-25] Such behavior is displayed in Fig. 1 where double superconducting transitions for two polycrystalline samples used in this work are shown. The temperature $T_{Ci}$ is related to the onset critical temperature at which superconducting clusters start to form, while $T_{Cj}$, defined at the branching point, is the temperature below which the superconducting clusters are connected via the Josephson effect which reduces the electrical resistivity to zero if a bias current is applied.[22-24]

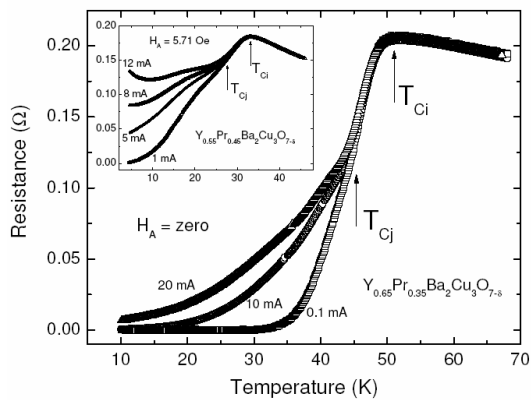

FIG. 1 – Double superconducting transitions for two granular samples. $T_{Ci}$ is the onset critical temperature and $T_{Cj}$ is the temperature below which superconducting clusters are connected via the Josephson effect.

Another important aspect to be considered in determining whether dissipation and weak coupling effects are related is the observation of hysteresis loops in magneto-resistance measurements.[25,26] If the loop exhibits a clockwise hysteresis, the dissipation is related to intergranular coupling.[25,26] On the other hand, if the loop has opposite direction, dissipation results from the motion of Abrikosov vortices.[26] In order to observe dissipation due to intragranular and intergranular effects in granular high-$T_C$ superconductors we measured magneto-resistance hysteresis loops under low and high magnetic fields. Results reported previously[27] show that the $R(H)$ curves display plateaus (ohmic regime) at intermediate magnetic field which can be used as a criterion to separate both intergranular and intragranular dissipation mechanisms.[27] $I-V$ curves measured below this ohmic regime show another important feature. The differential resistances in the linear regime of the $I-V$ curves are clearly magnetic field independent. We stress that this behavior cannot be explained by using the classical flux-flow model.[27-29] In order to explain these experimental observations we discuss the results within the framework of the RSJ model which predicts the existence of two-fluid particles, normal and superconducting electrons, carrying the total current[19-22] by taking into account a statistical theory based on similar aspects of the two-level systems.[30,31] The statistical model is applied to transport properties measurements performed on four granular high-$T_C$ superconducting samples, including hole-doped and electron-doped compounds.

## II. TWO-FLUID MODEL

In the two-fluid model suggested below it is supposed that both applied magnetic field and electrical current are related to the weak coupling effects and that they are not high enough to reach the limits expected to break Cooper pairs (current density is smaller than depairing current) or induce Abrikosov vortices inside superconducting samples ($H$ is much smaller than intragranular lower critical field).

As well accepted, there are two carriers types, normal and superconducting electrons, coexisting in equilibrium at temperatures below $T_C$. At a fixed temperature, there is a specific density ratio between normal particles and superconducting electrons predicted by the *static* two-fluid model.[32] On the other hand, if electrical current is passed through a superconducting sample, a dynamical situation is created and an interplay between the numbers of normal and superconducting electrons must take place to carry the applied current. At very low applied current, the current is supposed to be carried only by superconducting electrons (normal electrons do not contribute to the conduction) and as a consequence no voltage is observed in the sample. On the other hand, according to the RSJ model, if Josephson junctions are subjected to a high applied current, a normal current can flow parallel to the supercurrent resulting in a non-ohmic dissipative regime ($V \neq 0$).[17-20] Thus, if $I_O$ is defined as a characteristic supercurrent of a granular superconducting sample (supposed to be a network of Josephson-junctions[17-19]) at a fixed temperature and applied magnetic field, when $I \ll I_O$ the sample is in a true superconducting zero-resistance state and the total current that crosses the sample is carried by superconducting electrons. If $I \sim I_O$ a normal current flows parallel to the supercurrent and the total current should be transported by a specific ratio between the number of normal ($N_N$) and superconducting ($N_S$) electrons which should depend on the ratio $I/I_O$. Furthermore, in the $I \gg I_O$ limit, an ohmic regime should be reached. Following this idea, we suggest that there is a statistical ratio between $N_N$ and $N_S$ which is assumed to have similar physical aspects to those described by a simple two-level system.[30,31] The dynamical ratio between $N_N/N$ and $N_S/N$ at a fixed temperature could be written as:

$$N_N/N = e^{-\Delta I/I}/(e^{-\Delta I/I} + e^{\Delta I/I}), \quad (1)$$
$$N_S/N = e^{\Delta I/I}/(e^{-\Delta I/I} + e^{\Delta I/I}), \text{ and} \quad (2)$$
$$N_N + N_S = N \quad (3)$$

where $\Delta I/I \equiv (I_O - I)/I$ and $N$ is the total number of carriers transporting the current at a fixed temperature and under a constant applied magnetic field.

A graphical description of the statistical model is provided in Fig. 2 where are shown the fractional populations $N_N/N$ and $N_S/N$ as a function of $I/I_O$. The curves are symmetrical and their behaviors agree very well with the aspects expected by the RSJ model. If $I/I_O$ goes to zero the applied current is only carried by superconducting electrons ($N_S/N = 1$) and no normal electron crosses the sample ($N_N/N = 0$). At $I$ comparable to $I_O$, both normal and superconducting electrons carry the applied current, producing a voltage in the superconducting sample. Finally, at $I = I_O$, we obtain a clear definition of the characteristic current $I_O$ as the current at which the number of normal electrons carrying the applied current is equal to the number of superconducting electrons.

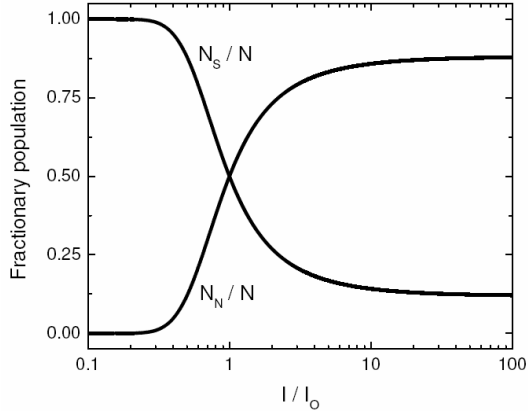

FIG. 2 – Fractional population of normal ($N_N/N$) and superconducting electrons ($N_S/N$) based on a two-level system model as a function of the normalized applied electrical current.

To determine the $N_N(I)$ and $N_S(I)$ dependencies it is necessary to find the behavior of $N(I)$. After a careful inspection of the Eq. (1) to (3), it is clear that $N$ is current dependent. For example, at $I \ll I_O$, $N = N_S$ and if $I$ is increased, $N$ increases. Thus, such as $N_N$ increases with decreasing $N_S$, we can suppose there exists a function such that $f(N_N) + f(N_S) =$ constant. In this work we have used the simplest form for such a function which is given by

$$N_N^{1/\beta} + N_S^{1/\beta} \equiv N_O^{1/\beta}, \qquad (4)$$

where $N_O$ is a constant and $\beta$ is neither zero nor 1 which can be determined by using experimental data.

Remembering that in the two-level system $N_N$ and $N_S$ are proportional to $e^{-\Delta I/I}$ and $e^{\Delta I/I}$, respectively, the Eq. (4) implies that the number of carriers must have the following current dependencies:

$$N_N = N_O/(e^{2\Delta I/\beta I}+1)^\beta \text{ and} \qquad (5)$$
$$N_S = N_O/(e^{-2\Delta I/\beta I}+1)^\beta. \qquad (6)$$

Since the dissipation is related to the number of normal electrons crossing the sample, we suppose that the voltage in a superconducting sample must be proportional to the normal electrons ratio ($N_N/N_O$) carrying part of the total current in the dissipative regime, i. e. $V = \alpha(N_N/N_O)R_N I$, where $R_N$ is the normal state shunt resistance which must be magnetic field independent.[18] Thus, the dissipation voltage in a superconducting sample should be given by

$$V = \alpha R_N I/(e^{2\Delta I/\beta I}+1)^\beta, \qquad (7)$$

where $\alpha = (e^{-2/\beta}+1)^\beta$ due to the fact that if $I_O$ vanishes an ohmic regime (normal state shunt resistance) should be reached ($V = R_N I$).

To compare our two-fluid model with experimental results, transport measurements performed on polycrystalline superconducting samples of the hole-doped $Y_{1-x}Pr_xBa_2Cu_3O_{7-\delta}$ and electron-doped $Sm_{2-x}Ce_xCuO_{4-\delta}$ systems are shown. The samples were prepared by the conventional solid state reaction. Details about preparation, characterization and superconducting properties have been reported elsewhere.[27-29,33] Essentially, all samples have double superconducting transitions as displayed in Fig. 1. In the ranges of magnetic field and current used during the measurements, the magneto-resistance curves of all samples displayed clockwise hysteresis loops, unambiguously demonstrating that dissipation effects are related to weak coupling mechanisms.

At $I \gg I_O$, it is possible to show from Eq. (7) that, independent of $\beta$, $dV/dI$ must approach the $R_N$ value, suggesting that $I-V$ curves should display parallel linear regimes at high applied currents if the shunt resistance is magnetic field independent, which is in agreement to the RSJ model.[17-20] Fig. 3 displays $dV/dI$ versus $I$ curves for the $Y_{0.55}Pr_{0.45}Ba_2Cu_3O_{7-\delta}$ sample. These measurements were carried out at 4.2 K under low applied magnetic fields after zero-field cooling by using a copper solenoid. We can see that the slopes approach a constant value at high current limit representing magnetic independent linear regimes in the $I-V$ curves which is in excellent agreement with the prediction of the two-fluid model reported here. Thus, we can suppose that the slope at high current limit is related to the normal state shunt resistance providing $R_N = 0.355\,\Omega$ at 4.2 K for the $Y_{0.55}Pr_{0.45}Ba_2Cu_3O_{7-\delta}$ sample.

Another way to determine $R_N$ is increase applied magnetic field up to the upper limit for weak coupling effects.[27] In such a case, $I_O$ should vanish and the magneto-resistance ($V/I$) will approach an ohmic regime with resistance equal to $R_N$. In the inset of the Fig. 3, the magneto-resistance for the $Y_{0.55}Pr_{0.45}Ba_2Cu_3O_{7-\delta}$ sample at 4.2 K measured under different applied currents is shown. As expected, the sample reaches ohmic behavior at high magnetic fields with the same shunt resistance value obtained from the $I-V$ curves.

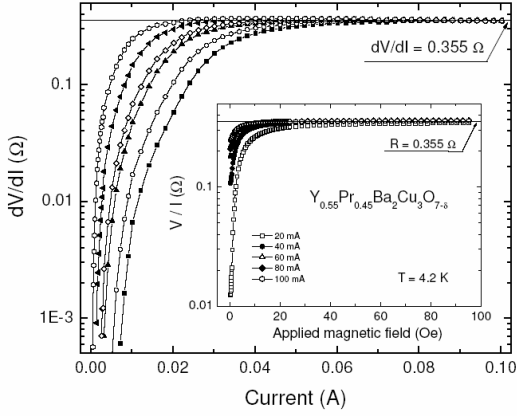

FIG. 3 – $dV/dI$ calculated from some $I-V$ curves measured in the $Y_{0.55}Pr_{0.45}Ba_2Cu_3O_{7-\delta}$ sample at 4.2 K under different applied magnetic fields (from right: $H$ = zero, 0.95, 1.43, 1.90, 2.85, and 5.71 Oe) as function of the current. In inset is presented the magneto-resistance behavior under different applied currents. The saturation regimes at $dV/dI$ and $V/I = 0.355\,\Omega$ are in agreement with the predictions of the two-fluid model.

In order to obtain the $\beta$ value, we note that the plot $V/I^2$ against $I$ must display peaks at $I=I_O$, independently of $\beta$. On the other hand, the magnitude of the peak depends on $\beta$ as

$$V/I^2(I=I_O) = R_N[(e^{-2/\beta}+1)/2]^\beta /I \qquad (10)$$

which can be obtained from careful inspection of the experimental data. In Fig. 4(a) $V/I^2$ versus $I$ for the $Y_{0.55}Pr_{0.45}Ba_2Cu_3O_{7-\delta}$ sample is shown. All curves reveal peaks in good agreement with the prediction of the Eq. (10). Fig. 4(b) highlights the region near the peaks for the three lowest applied magnetic fields. The behavior expected for three different $\beta$ values using $R_N = 0.355\,\Omega$ are also plotted. It is evident that the magnitude of the peaks are well described by $\beta \sim 2$.

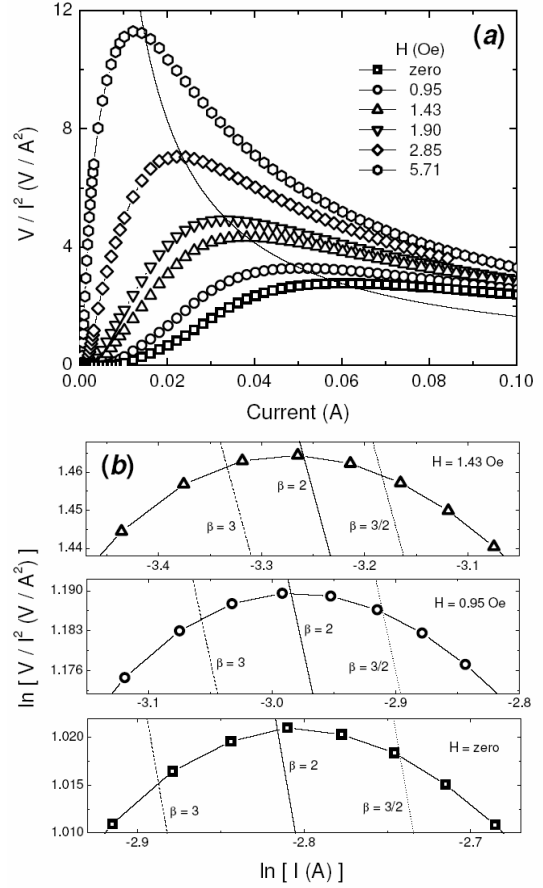

FIG. 4 – (a) $V/I^2$ versus $I$ measured in the $Y_{0.55}Pr_{0.45}Ba_2Cu_3O_{7-\delta}$ sample at 4.2 K under different applied magnetic fields. The solid line represents the expected behavior of the peaks using Eq. 10 with $\beta=2$. In (b) are displayed three curves blowed up near of the peaks. The lines are the behaviors of the peaks for different $\beta$ values at $I=I_O$.

Another important point of Fig. 4 is that the $I_O(H)$ values can be determined directly from the curves by finding the points where $d(V/I^2)/dI = 0$. It is observed that $I_O$ decreases with increasing applied magnetic field, in agreement with the assumption that the coupling effects deteriorate if magnetic field is increased in the intergranular region.

By using the Eq. (7) with $\beta=2$, it is easy to show additional implications can arise naturally from the two-fluid model. For example, in the $I \ll I_O$ limit ($N_N = 0$), $V$ vanishes in agreement with the $I-V$ characteristic curves. Furthermore, if $I = I_O$, $N_N = N_S$ providing $I_N \sim I_S \sim I_O/2$ which is in good agreement with our previous description[27-29] of transport properties by the RSJ model ($I = I_S + I_N$ and $V/I = R_N$).[18] On this issue, we should observe that $V = 1.871 R_N I_O /4$ and $dV/dI = 1.871 R_N / 2$ which can be carefully compared with $I-V$ characteristic curves. The results for $V$ at $I = I_O$ are plotted as a function of $I_O$ in the inset of the Fig. 5a. $V$ is proportional to $I_O$ in good agreement with the expected slope (see solid line). Also, the mean value for $dV/dI$ at $I = I_O$ is $0.325 \pm 0.011$, which is very close to the expected value ($0.332$) if we use $R_N = 0.355 \, \Omega$.

By using $R_N = 0.355 \, \Omega$ and $I_O(H)$ values, we are able to reproduce the experimental $I-V$ curves without fitting parameters. Fig. 5(a) displays some experimental curves (symbols) for the $Y_{0.55}Pr_{0.45}Ba_2Cu_3O_{7-\delta}$ sample plotted together with solid lines calculated with Eq. (7) and $\beta = 2$. In Fig. 5(b) the same data are plotted in a collapsed curve. These results demonstrate that the experimental curves are in excellent agreement to the two-fluid model proposed here.

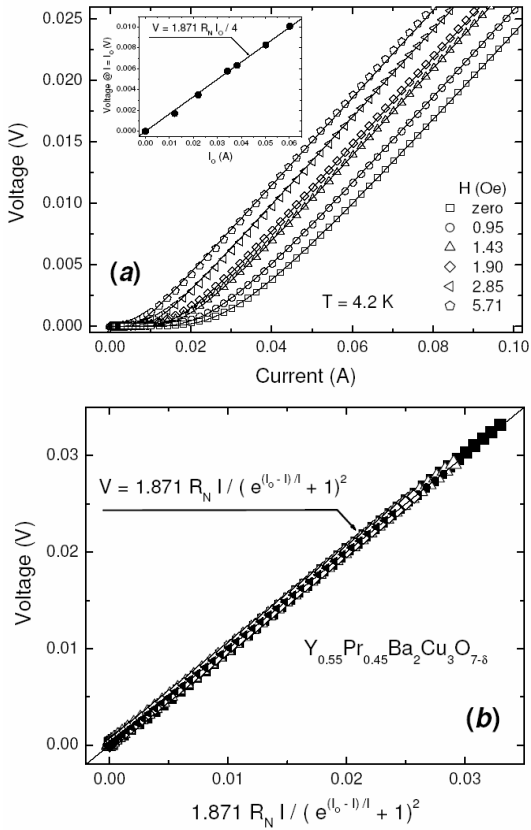

FIG. 5 – (a) Experimental (symbols) and calculated (solid lines) $I-V$ characteristic curves using $R_N = 0.355 \, \Omega$ and $I_O(H)$ values obtained from Fig. 4 by taking $d(V/I^2)/dI = 0$. In (b) are shown the data plotted in a collapsed curve based on equation (7) with $\beta = 2$. In inset is displayed the voltage values at $I = I_O$ point (the solid line is the behavior expected by the two-fluid model).

Finally, in order to verify that the two-fluid model can be applied to other superconducting samples. In Fig. 6 is presented the $V/I^2$ values at the peak normalized by $R_N$ of each sample versus $1/I_O$. The experimental results collapse very close to the theoretical curve over more than one order of magnitude, suggesting that the model proposed here is independent of the sample composition, critical temperature as well as whether or not the superconducting compound is electron or hole-doped.

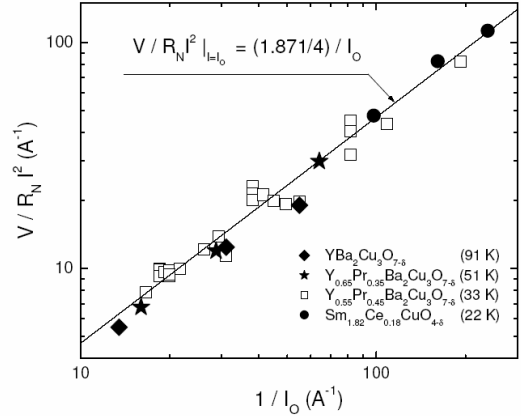

FIG. 6 – Peak of $V/R_N I^2$ at $I = I_O$ versus $1/I_O$ for some granular superconducting samples (symbols). The solid line represents the expected behavior predicted by two-fluid model reported here. The temperatures indicated in the parentheses are the $T_{Ci}$ of the samples.

### III. CONCLUSION

This work proposes a two-fluid model to describe transport properties of granular superconductors. The model takes into account a statistical ratio between the number of normal and superconducting electrons carrying the applied current. Several implications of the model are obtained which agree very well with transport properties of different high-$T_C$ superconductors. It is observed that only two parameters ($R_N$ and $I_O$), directly obtained from experimental curves, are necessary to describe $I-V$ curves quantitatively without fitting parameters. The discussion of the results obtained in different superconducting compounds suggest that the two-fluid model is independent of the sample composition, critical temperature and whether the compound is electron or hole-doped.

## ACKNOWLEDGEMENTS

The authors are grateful to R. F. Jardim for providing some samples of the $Sm_{2-x}Ce_xCuO_{4-\delta}$ system. J. J. Neumeier is acknowledged for his valuable comments regarding the final version of this paper. Authors also thank C. Y. Shigue and C. Bormio-Nunes for many stimulating discussions. This work was supported by FAPESP (97/11113-6 and 00/03610-4) and CAPES.